\documentclass[aps,prd,sh
owpacs,showkeys]{revtex4}
\begin{document} 
\title{Comments on a bound state model for a two body system}  
\author{L.Micu}
\email{lmicu@theory.nipne.ro}
\affiliation{Department of
Theoretical Physics\\ Horia Hulubei Institute for Physics and Nuclear
Engineering, Bucharest POB MG-6, RO 077125}
\date{\today}

\begin{abstract}
We show that in classical mechanics, as well as in nonrelativistic quantum 
mechanics the equation of the relative motion for a two-body bound system 
at rest can be replaced by individual dynamical equations of the same kind as the first one, but with different parameters. We assume that in relativistic quantum mechanics the individual equations are Dirac equations with modified  parameters in agreement with the individual Schr\"odinger equations.   
We find that products of solutions to the individual equations with correlated arguments are the quantum analogues of the classical representation of a bound system and represent suitable models for the bound state wave functions. 
As validity test for the new representation of bound states we suggest to use some observable differences between this one and the representation in terms of relative coordinates.  
\end{abstract}



\maketitle

\section{Introduction}

In the textbooks for classical mechanics and nonrelativistic quantum mechanics (see for instance \cite{cm} and \cite{qm}) the approaches to the two body problem start with the replacement of the dynamical equation by two equations: the first one describes the free motion of the center of mass while the second equation describes a particle with the mass equal to the reduced mass, moving in an external potential well. A long experience has shown that the solution of the last one offers a suitbale representation of the bound state in the case of the heavy-light systems at rest, where it may be assumed that the position of the heavy particle coincides with that of the center of mass and the relative motion may be seen like the motion of the light particle with respect to the center of mass. 

In all the other cases the relative position vector is not sufficient for an adequate representation of a bound system. Moreover, the separation of the center of mass from the relative coordinates leaves the intrinsic properties of the particles, like spin or charges, outside the dynamical scheme, because these ones cannot be associated to either of them. Therefore, whenever the intrinsic properties play a significant r\"ole, the bound system must be represented in terms of the individual coordinates with respect to the center of mass which is supposed fixed. This is particularly important when the internal particles are individualized by their specific behaviour with respect to an external force. This is for instance the case of semileptonic decays of the heavy-light mesons where only the heavy quark is decaying, the other being spectator.

In this paper we present a procedure of introducing the individual coordinates and equations of motion in the case of a bound system made of two bodies with arbitrary masses. We start from the observation that in the rest frame of the bound system the relative position vector and the individual position vectors of the internal bodies with respect to the center of mass are collinear. Using this fact we transform the equation of the relative motion into individual equations of the same kind as the first one but with different parameters and express the solution to the bound state problem in terms of individual solutions.

In the next section the procedure is applied to nonrelativistic classical systems. We recall the well known fact that the representation of the bound system in terms of the relative coordinates is equivalent with the representation in terms of correlated individual coordinates. 

In the third section the procedure is extended to nonrelativistic quantum systems. From the Schr\"odinger equation for the relative motion we derive similar individual equations with modified parameters and write the complete wave function of the bound system as a product of individual wave functions and of the localized solution of the free wave equation describing the motion of the center of mass. The procedure gives rise to some specific effects which can be used as tests of validity.

The case of relativistic quantum systems is discussed in the fourth section. Recalling that relativistic covariance is incompatible with the separation of the relative and individual coordinates we consider that the Dirac equation with interaction potential must be seen as an individual equation. The parameters are chosen in agreement with the individual Schr\"odinger equation of the kind defined in the previous section. As a test of validity we calculate the mass and the energy levels of the hydrogen atom and notice the coincidence with the values obtained with the usual procedure up to terms of order $\alpha^4$.

In the last section we comment briefly on the relation between the coordinate and momentum representations of a bound system.

Throughout the paper we use the following notations:
$\vec{r}_1,~\vec{r}_2$ and $\vec{r}=\vec{r}_1-\vec{r}_2$ are the individual and relative position vectors in an arbitrary coordinate system.  The center of mass position vector is defined as: 
\begin{equation}\label{id1}
\vec{r}_{CM}=\eta_1\vec{r}_1+\eta_2\vec{r}_2
\end{equation}
where $\eta_{1,2}=\frac{m_{1,2}}{m_1+m_2}$ with $m_{1,2}$ the masses of the internal particles. Obviously,
\begin{equation}
\pm\vec{r}_{1,2}=\vec{r}_{CM}+\eta_{2,1}\vec{r}
\end{equation}
and hence, whenever the center of mass coincides with the origin of the coordinate system one has
\begin{equation}\label{id2}
\vec{r}={\vec{r}_1\over\eta_2}=-{\vec{r}_2\over\eta_1}.
\end{equation}

\section{Classical mechanics}

In classical mechanics the Newton equation of motion for the particle with reduced mass is:
\begin{equation}\label{eqN}
\eta_1\eta_2(m_1+m_2){d^2\vec{r}(t)\over 
dt^2}=-\vec{\nabla}_{r}V\left(r(t)\right)\,
\end{equation}
where $V(r)$ represents the interaction potential. 

Using the relations (\ref{id1}) and (\ref{id2}) which are valid in the 
center of mass frame in the case of an isolated system even when 
$\vec{r},~\vec{r}_1,~\vec{r}_2$ are functions of $t$, we observe that the 
equation (\ref{eqN}) can be written in either one of the two forms:
\begin{equation}\label{eqN1}
m_1{d^2\vec{r}_1(t)\over dt^2}=-\eta_2\vec{\nabla}_{r_1}V(\eta_2^{-1}r_1)
\end{equation}
or
\begin{equation}\label{eqN2}
m_2{d^2\vec{r}_2(t)\over dt^2} 
=-\eta_1\vec{\nabla}_{r_2}V(\eta_1^{-1}r_2).
\end{equation}
We note that in the heavy-light case, where $\eta_2\approx0$ the heavy 
particle is at rest and the light one turns around it, as expected. 

The solutions to (\ref{eqN1}) and (\ref{eqN2}) describe the individual motion of the physical particles with the masses $m_1$ and $m_2$ in the external potentials $\eta_2V(\eta_2^{-1}r_1)$ and  $\eta_1V(\eta_1^{-1}r_2)$ which are of the same kind as $V(r)$. If $\vec{r}_{CM}=0$ one can say that the instantaneous positions are at the ends of a segment of variable length, turning around the center of mass which is fixed and divides the segment in two parts in the ratio $-\eta_2\eta_1^{-1}$. Noticing the similarity of equations of motion (\ref{eqN1}) and (\ref{eqN2}) we also can say that the trajectories are similar and the instantaneous momenta of the two particles are equal and opposite. 
 
Now resorting to (\ref{id2}) it can be easily shown that the sum of the total 
energies of the individual particles is equal with the energy of the 
nonphysical particle with reduced mass. One finds indeed:
\begin{eqnarray}
&&E_1={m_1\over2}\left({d\vec{r}_1(t)\over 
dt}\right)^2+\eta_2V\left(\eta_2^{-1}r_1(t)\right)=\eta_2E\label{e1}\\
&&E_2={m_2\over2}
\left({d\vec{r}_2(t)\over 
dt}\right)^2+\eta_1V\left(\eta_1^{-1} r_2(t)\right)=\eta_1E\label{e2}
\end{eqnarray}
where
\begin{equation}\label{e}
E={\eta_1\eta_2(m_1+m_2)\over 2}~\left({d\vec{r}(t)\over 
dt}\right)^2+V(r(t)).
\end{equation}
Similar relations hold for the angular momenta and hence one may conclude that in classical physics a bound system at rest may be represented either in terms of the relative coordinates or of the individual coordinates with the constraint (\ref{id2}). 

\section{Nonrelativistic quantum mechanics}

In quantum mechanics the internal state of a bound system is supposed to be described by a solution with finite norm, $\psi(\vec{r})$, of the time independent Schr\"odinger equation having the generic form:  
\begin{equation}\label{eqS}
({\mathcal H}-E) \psi(\vec{r})=0
\end{equation}
where 
\begin{equation}\label{hS}
{\mathcal H}(r)=-\frac{1}{2~\eta_1\eta_2(m_1+m_2)}\vec{\nabla}^2_{r}+V(\vec{r}).
\end{equation}
$E$ is the eigenvalue of $\mathcal H$ corresponding to $\psi$ and represents 
the quantized value of the internal energy of the bound system which is the sum of the kinetic and potential energies of 
a particle with reduced mass in the external field $V(r)$. 

Like in classical mechanics, by using (\ref{id2}), the dynamical 
equation for the relative motion (\ref{eqS}) of spinless particles can be written in terms of the individual coordinates either as: 
\begin{equation}\label{eq1}
\left(-\frac{\eta_2^2}{2m_1}~\vec{\nabla}^2_{r_1}+
\eta_2~V(\eta_2^{-1}r_1)-\eta_2E\right)\psi_{n1}(\vec{r}_1)=0
\end{equation}
or as
\begin{equation}\label{eq2}
\left(-\frac{\eta_1^2}{2m_2}~\vec{\nabla}^2_{r_2}+
\eta_1~V(\eta_1^{-1}r_2)-\eta_1E) \psi_{n2}(\vec{r}_2\right)=0.
\end{equation}
Obviously, the equations (\ref{eqS}), (\ref{eq1}) and (\ref{eq2}) with the same kind of boundary conditions are nothing else but different ways of writing the same Schr\"odinger equation. However, it is worth noticing that the solutions to the last two equations, $\psi_{n_1}$ and $\psi_{n_2}$, represent two physical particles with masses $m_1$ and $m_2$ individually bound in the attractive external potential wells $\eta_2~V(\eta_2^{-1}r_1)$ and  $\eta_1~V(\eta_1^{-1}r_2)$ respectively. (We remark that their distance to the corresponding centers of forces are $\eta_2^{-1}r_1=\eta_1^{-1}r_2=r$.) In view of the agreement with the classical case where the individual orbits are similar, we assume that the wave functions $\psi(\vec{r}),~\psi_{n_1}(\vec{r}_1)$ and $\psi_{n_2}(\vec{r}_2)$ are similar functions of coordinates which means that their spatial quantum numbers are equal. Also, as it results from (\ref{eq1}) and (\ref{eq2}), the corresponding energy eigenvalues are $E,~E_1=\eta_2E$ and $E_2=\eta_1E$. 

We notice that the difference between $\psi,~\psi_{n_1},$ and $\psi_{n_2}$ comes from the different values of the parameters in the interaction potential and from the different units of length used in the $\vec{r}$, $\vec{r}_1$, and $\vec{r}_2$ spaces. Like in classical case the units of length satisfy the relation: $u_r:u_{r_1}:u_{r_2}=1:\eta_2:\eta_1$. 

Now taking into account the individual character of the wave functions $\psi_{n_1}$ and $\psi_{n_2}$ we assert that the wave function of the bound system at rest can be written as follows:
\begin{equation}\label{Psi}
\Psi(\vec{r}_1,\vec{r}_2)=
\psi_{n_1}(\vec{r}_1-\vec{r}_{CM})\psi_{n_2}(\vec{r}_2-\vec{r}_{CM})~
\Delta_\sigma(\vec{r}_{CM})
\end{equation}
where the function $\Delta_\sigma$ is the solution of the Schr\"odinger equation for the center of mass which is localized in a region of range $\sigma^{-1}$ around $\vec{r}_{CM}=0$. In the limit $\sigma\to\infty$,  $\Psi^2_{n,\nu_1,\nu_2}$ is the probability density to find the two particles at the ends of a segment of length $\vert(\vec{r}_1-\vec{r}_2)\vert$ divided by the center of mass in two parts in the ratio $-\frac{\eta_2}{\eta_1}$. 

We remark that $\Psi_{n,~\nu_1,\nu_2}(\vec{r}_1,\vec{r}_2)$ satisfies the main conditions to be considered a suitable model for the bound state wave function: its $L^2$ norm in the $R^3\times R^3$ space is finite and in the limit $\sigma\to \infty$ the representation of the bound system in the coordinate space is in agreement with the classical representation.

However, we notice that, unlike the classical case, where the correlation of the individual positions in the rest frame entails the correlation of the individual momenta, in the quantum case the perfect correlation of the individual positions transforms into a total lack of correlation of the individual momenta. Indeed, if we take $\Delta_\sigma(\vec{r}_{CM})=\left(\frac{\sigma}{\sqrt{2\pi}}\right)^\frac{3}{2}{\mathrm exp}[-\vec{r}_{CM}^2\sigma^2]$ and then expand the individual wave functions in terms of momentum eigenstates and integrate over the degrees of freedom of the center of mass we find 
\begin{equation}\label{cm}
\int d^3r_{CM}\Psi(\vec{r}_1,\vec{r}_2)
=(2\pi)^3\int d^3k_1 d^3k_2 d^3Q
\tilde{\psi}_{n_1}(\vec{k}_1){\mathrm e}^{-i\vec{k}_1\vec{r}_1}\tilde{\psi}_{n_2}(\vec{k}_2){\mathrm e}^{-i\vec{k}_2\vec{r}_2}\tilde{\Delta}(\vec{Q})~\delta^{(3)}(\vec{k}_1+\vec{k}_2-\vec{Q})
\end{equation}
where $\tilde{\psi}(\vec{k})=(2\pi)^{-3}\int d^3r \psi(\vec{r}){\mathrm e}^{i\vec{k}\vec{r}}$ and $\tilde{\Delta}=\left(\sqrt{2\pi}\sigma\right)^{-\frac{3}{2}}{\mathrm exp}[-{1\over4\sigma^2}(\vec{p}_1+\vec{p}_2)^2]$. It follows that the perfect correlation of the individual positions ($\vec{r}_{CM}=0$) is achieved for $\sigma\to\infty$, while the perfect correlation of the individual momenta ($\vec{k}_1+\vec{k}_2=0$) results for $\sigma\to0$. 

Then in order to have a certain degree of correlation both of the individual positions and of the individual momenta the parameter $\sigma$ must be finite.
This means that the center of mass of the two particles is distributed in a region of range $\sigma^{-1}$ around the origin or, in other terms, it is not fixed when the bound system is at rest.
This might appear as a misleading view but it could be accepted if we resort to the quantum field representation of a bound system where the binding forces are generated by the continuous exchange of quanta between the constituents. Then the motion of the center of mass may be seen as the consequence of the imperfect cancellation of momenta during the quantum fluctuations which generate the binding.

Closing this section we notice that in the new representation of a bound system there are two individual wave functions instead of a single function of the relative coordinates. Accordingly, a term in the relative wave function behaving like $r^l$ when $r\to 0$ is replaced by $r^{2l}$ in the limit $\sigma\to\infty$. Also, if the angular momentum had the eigen value  $l$ in the old representation, now it may have any value in the range $[0,2l]$. Noticing the direct influence of these changes on the form factors we consider that a comparison of the calculated values with the experimental data may provide a good test of validity for the present model.

\section{ Relativistic quantum mechanics}

As it is well known the separation of the center of mass coordinates from the relative ones cannot be performed in the case of the Dirac equation and hence we are prevented from applying the procedure developed in the preceding section to fermion systems. 

In exchange, we are allowed to assume that the Dirac equation with an interaction potential is one of the individual equations if it is in agreement with one of the Schr\"odinger equations written above. Then, taking (\ref{eq1}) as a model, the suitable form of the individual Dirac equation for the particle {\it 1} is the following:
\begin{equation}\label{eqd}
(-i\eta_2\vec{\nabla}_{r_1}\vec{\alpha}_1+\beta_1 m_1+\eta_2V(\eta_2^{-1}r_1)-E_1)\psi_{n_1}(\vec{r}_1)=0.
\end{equation}
The second individual equation may be obtained from (\ref{eqd}) by interchanging the indices 1 and 2.

The wave function of the bound fermion system is supposed to be of the same form like (\ref{Psi}) where this time the individual wave functions $\psi_{n_{1,2}}$ are bi-spinors and $\Delta$ is simply a function constraining the center of mass degrees of freedom in agreement with the nonrelativistic case. 
The mass of the bound system is the total energy of the bound system in its rest frame and hence it is the sum $E_1+E_2$ of the eigenvalues of the Dirac Hamiltonian which include the individual rest masses. 

A simple test of validity of the present model is to compare the difference $\mathcal D$ between the energy levels of a hydrogen-like atom using the new and the old prescriptions. Specifically, one has to calculate the difference between the energy of the bound system at rest written as the sum of the individual energies of the electron and of the nucleus in Coulomb potentials with modified strengths as it results from (\ref{eqd}) and the energy obtained by solving the Dirac equation for a particle with reduced mass in the Coulomb potential. 

Considering the expressions quoted in Ref.\cite{iz} for the energy levels, in the first case we have 
\begin{equation}\label{mass}
M_a=E_1+E_2
\end{equation}
where 
\begin{equation}\label{e12}
E_{1,2}=m_{1,2}~\left[1+\frac{(\eta_{2,1} Z\alpha)^2}{(n-\delta^{j}_{1,2})^2}\right]^{-\frac{1}{2}}
\end{equation}
with 
\begin{equation}
\delta^{j}_{1,2}=j+\frac{1}{2}-\left[\left(j+\frac{1}{2}\right)^2-
(\eta_{2,1}Z\alpha)^2\right]^{\frac{1}{2}}.
\end{equation}

The energy of the particle with reduced mass in the Coulomb field of the nucleus is obtained from $E_1$ by replacing $\eta_2\to 1$ and $m_1\to
\frac{m_1 m_2}{(m_1+m_2)}$. Then, after including the recoil corrections (see \cite{iz}), the difference between the values of the energy levels becomes: 
\begin{equation}\label{delta}
{\mathcal D}\approx\frac{Z^4\alpha^4(m_1+m_2)}{n^3}
\left\{ \frac{3}{8n}\left[1-\frac{m_1m_2}{(m_1+m_2)^2}\left(1-\frac{m_1~m_2}{3(m_1+m_2)^2}\right)\right]-\frac{1}{2j+1}
\left[1-\frac{m_1~m_2}{(m_1+m_2)^2}\right]\right\}
\end{equation}
where $Z$ is the nucleus charge. The terms in the first square brackets in (\ref{delta}) represent the difference between the recoil corrections and can be measured in principle. We mention that in the old treatment the recoil effect has been obtained by using some particular assumptions, while in the present formalism it results from the individual treatment of the particles. The last part is the difference between the fine splitting terms which is negligible small in the case of hydrogen-like atoms where $\frac{m_2}{m_1}<<1$.
 
Closing this section we notice an important consequence of the individual, relativistic treatment of the internal particles: the attractive forces create  a positive mass defect $\delta_M=(m_1+m_2)-M_a$ which tends to 0 when $Z\alpha\to0$ and is automatically included into the mass of the bound system together with the free masses. From the formal point of view this is perhaps the most impressive result of the present model. 

\section{Comments and conclusions}

In the last two sections we have shown that the procedure leading to a representation of a bound system in terms of individual coordinates outlined in classical mechanics also may be extended to quantum physics with similar results. 

However, it is worth noticing some of the pecularities of the quantum representations. A first one is that in quantum mechanics the perfect correlation of the individual coordinates is incompatible with the perfect correlation of the individual momenta and conversely. This is unlike the classical case where the correlation of the individual position entails the correlation of the individual momenta.

A second pecularity concerns the dependence on time of the bound state wave function in the coordinate and in the momentum representation. In the first case the wave function is stationary and its dependence on time concentrates in the overall factor ${\mathrm exp}[i(E_1+E_2)t]$. In the momentum space the bound system is represented by the projections of the wave function on two particle free states whose dependence on time is ${\mathrm exp}[i(e_1+e_2)t]$ where $e_{1,2}$ are the relativistic free energies. Observing that in the relativistic case $E_1+E_2=M$, while $e_1+e_2$ can take any value in the range $[(m_1+m_2),\infty]$ one may conclude immediately that a two body bound system cannot be adequately represented in terms of two body free states. This is not unexpected, because neither the  potential energy of the bound system, nor the correlation of the individual coordinates can be explained with their aid. In principle this ones ought to be explained in field theory but, if one wishes to obtain a suitable representation of a bound system in the framework of the quantum mechanics, it is convenient to accept the existence of a neutral, vacuum-like, effective constituent beside the "valence" particles \cite{micu}, whose features compensate the differences between the momentum and the coordinate representations. Specifically, the momentum of this one must compensate the free individual momenta and its contribution to the energy must supply the system with the potential energy which misses in the momentum representation.

Obviously, the effective constituent is not an elementary excitation of the background field and it is hard to imagine its representation in the coordinate space. However, taking into account its r\^ole in the bound system, it may be associated with the external potential which generates the forces binding the valence particles together. Then, recalling that in field theories binding is the result of the continuous exchange of quanta, we consider reasonable to assume that the effective constituent is an average representation of countless quantum fluctuations.

\begin{acknowledgments}
The work was finished during author's visit at the 
Center of Theoretical Physics in Marseille in the frame of 
the Cooperation Agreement between CNRS and the Romanian 
Academy. The hospitality at the Center of Theoretical Physics 
and the clarifying discussions with Dr. Claude Bourrely are warmly acknowledged. 
Special thanks are due to Fl. Stancu for valuable suggestions and continuous encouragement and to I. Caprini and P.Dita for useful comments. 
The financial support from the Ministry of Education and 
Research in the frame of the CERES Programme under the Contract No. 4-123/2004 is acknowledged.
\end{acknowledgments}

\end{document}